 \documentclass[final,authoryear,times,twocolumn]{elsarticle}

\usepackage{amssymb}

\biboptions{sort&compress}

\journal{Annals of Nuclear Energy}

\begin{document}

\begin{frontmatter}



\title{Activation cross-sections of longer-lived products of proton induced nuclear reactions on dysprosium up to 36 MeV}


\author[1]{F. T\'ark\'anyi}
\author[1]{F. Ditr\'oi\corref{*}}
\author[1]{S. Tak\'acs}
\author[2]{A. Hermanne}
\author[3]{A.V. Ignatyuk}
\cortext[*]{Corresponding author: ditroi@atomki.hu}

\address[1]{Institute for Nuclear Research of the Hungarian Academy of Sciences (ATOMKI),  Debrecen, Hungary}
\address[2]{Cyclotron Laboratory, Vrije Universiteit Brussel (VUB), Brussels, Belgium}
\address[3]{Institute of Physics and Power Engineering (IPPE), Obninsk, Russia}

\begin{abstract}
Activation cross-sections of longer-lived products of proton induced nuclear reactions on dysprosium were measured up to 36 MeV by using stacked foil irradiation technique and  $\gamma$-spectrometry. We report for the first time experimental cross-sections for the formation of the radionuclides $^{162m}Ho$, $^{161}Ho$, $^{159}Ho$, $^{159}Dy$, $^{157}Dy$, $^{155}Dy$, $^{161}Tb$, $^{160}Tb$, $^{156}Tb$ and $^{155}Tb$. The experimental data were compared with the results of cross-section calculations of the ALICE and EMPIRE nuclear model codes and of the TALYS nuclear reaction model code as listed in the on-line libraries TENDL-2011 and TENDL-2012. 
\end{abstract}


\end{frontmatter}


\section{Introduction}
\label{1}
A research program is running to study activation cross-sections of proton and deuteron induced reactions mainly for practical applications and to test the presently used theoretical nuclear reaction codes. From a detailed study of the literature it was recognized that data on rare earth elements in most cases are missing. It is well known that many rare earth radionuclides are used in medicine for diagnostic (PET) and, in a larger proportion, for therapeutic (radiopharmaceuticals and brachytherapy) purposes. 
In the frame of this systematic study we have investigated the activation cross-sections induced by protons and deuterons on natural dysprosium targets. The part of the study specifically devoted to production of $^{161}Ho$, a candidate therapeutic radioisotope, was published separately \citep{TF2013}. Here we report on the complete set of activation cross-section data induced by proton irradiation of dysprosium. No earlier experimental data were found in the literature.

\section{Experiment and data evaluation}
\label{2}
The general characteristics and procedures for irradiation, activity assessment and data evaluation (including estimation of uncertainties) were similar as in many of our earlier works \citep{TS,TF2012}.
The main experimental parameters and the methods of data evaluation for the present study are summarized in Table 1 \citep{Andersen, Bonardi, Canberra, Dityuk, Herman, Error, Kinsey, Koning, Pritychenko, SZG, TF1991, TF2001}. The used decay data are collected in Table 2.
For beam current and beam energy monitoring and for energy degradation Ti foils were incorporated downstream of each dysprosium foils in the stack. All monitor foil data were considered simultaneously in order to obtain the beam current and beam energy in each target foil by comparison with the IAEA recommended monitor data \citep{TF2001}. The measured cross-sections of the monitor reaction and the recommended data are shown in Fig. 1.

\begin{figure}[h]
\includegraphics[scale=0.3]{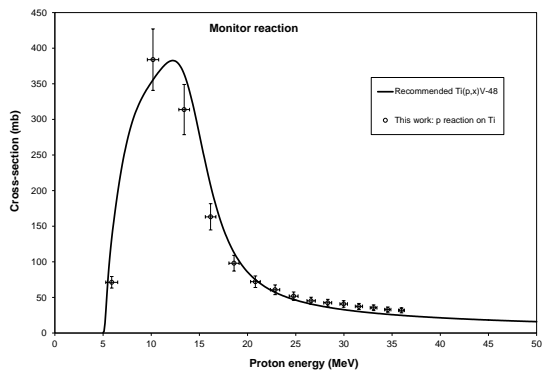}
\caption{The simultaneously measured monitor reactions for determination of proton beam energy and intensity}
\end{figure}

\begin{table*}[t]
\tiny
\caption{Main parameters of the experiment and the methods of data evaluations }
\centering
\begin{center}
\begin{tabular}{|p{1.1in}|p{1.1in}|p{1.0in}|p{1.0in}|} \hline 
\multicolumn{2}{|p{1in}|}{\textbf{Experiment}} & \multicolumn{2}{|p{2.1in}|}{\textbf{Data evaluation}} \\ \hline 
 &  &  &  \\ \hline 
Incident particle & Proton  & $\gamma$-spectra evaluation & Genie 2000, FORGAMMA \citep{Canberra,SZG} \\ \hline 
Method  & Stacked foil & Determination of beam intensity & Faraday cup (preliminary)\newline Fitted monitor reaction (final) \citep{TF1991} \\ \hline 
Target  stack and thicknesses  & Ti-Al-Dy-Al block\newline Repeated 15 times\newline ${}^{nat}$Ti foil, 10.9 mm\newline ${}^{nat}$Al foil, 98 mm\newline ${}^{nat}$Dy foil, 100.59 mm & Decay data & NUDAT 2.6 \citep{(Kinsey} \\ \hline 
Number of Dy target foils & 15 & Reaction Q-values & Q-value calculator \citep{Pritychenko} \\ \hline 
Accelerator & CGR 560 cyclotron Vrije Universiteit Brussels & Determination of  beam energy & \citep{Andersen} (preliminary)\newline Fitted monitor reaction (final) \citep{Andersen} \\ \hline 
Primary energy & 36 MeV & Uncertainty of energy & Cumulative effects of possible uncertainties \\ \hline 
Irradiation time & 71 min & Cross-sections & Isotopic cross-section \\ \hline 
Beam current & 61 nA & Uncertainty of cross-sections & Sum in quadrature of all individual contribution \citep{Error} \\ \hline 
Monitor reaction [recommended values]  & ${}^{nat}$Ti(p,x)${}^{48}$V reaction \citep{TF2001} & Yield & Physical yield \citep{Bonardi} \\ \hline 
Monitor target and thickness & ${}^{nat}$Ti, 10.9 mm & Theory & ALICE-IPPE \citep{Dityuk}\newline EMPIRE \citep{Herman}\newline TALYS (TENDL 2011, 2012 \citep{Koning} \\ \hline 
detector & HPGe &  &  \\ \hline 
$\gamma$-spectra measurements & 4 series &  &  \\ \hline 
Cooling times & 1.5 h, 20 h, 80 h, 330 day &  &  \\ \hline 

\hline
\end{tabular}
\end{center}
\end{table*}

\begin{table*}[t]
\tiny
\caption{Decay characteristics of the investigated reaction products and Q-values of reactions for their productions }
\centering
\begin{center}
\begin{tabular}{|p{0.6in}|p{0.5in}|p{0.6in}|p{0.5in}|p{0.9in}|p{0.8in}|} \hline 
\textbf{Nuclide} & \textbf{Half-life} & \textbf{E${}_{\gamma}$(keV)} & \textbf{I${}_{\gamma}$${}_{?}$(\%)} & \textbf{Contributing reaction} & \textbf{Q-value\newline (keV)} \\ \hline 
\textbf{${}^{162m}$Ho\newline }IT: 62 \%\newline $\varepsilon $: 38 \%\textbf{\newline } & 67.0 min & 57.74 & 4.4 & ${}^{162}$Dy(p,n)\newline ${}^{163}$Dy(p,2n)\newline ${}^{164}$Dy(p,3n) & -2922.04\newline -9193.05\newline -16851.18 \\ \hline 
\textbf{${}^{161}$Ho\newline }$\varepsilon $: 100 \%\textbf{} & 2.48 h & 77.42\newline 103.05\newline 157.26\newline 175.42 & 1.9\newline 103.05\newline 0.49\newline 0.43 & ${}^{161}$Dy(p,n)\newline ${}^{162}$Dy(p,2n)\newline ${}^{163}$Dy(p,3n)\newline ${}^{164}$Dy(p,4n) & -1640.64\newline -9837.63\newline -16108.65\newline -23766.77 \\ \hline 
\textbf{${}^{159}$Ho\newline }$\varepsilon $: 99.76 \%\newline ${\beta}^{+}$: 0.24 \% & 33.05 m & 121.012\newline 131.973\newline 252.963\newline 309.594\newline 838.625 & 36.2\newline 23.6\newline 13.7\newline 17.2\newline 3.84 & ${}^{160}$Dy(p,2n)\newline ${}^{161}$Dy(p,3n)\newline ${}^{162}$Dy(p,4n)\newline ${}^{163}$Dy(p,5n)\newline ${}^{164}$Dy(p,6n) & -11195.85\newline -17650.25\newline -25847.24\newline -32118.26\newline -39776.37 \\ \hline 
\textbf{${}^{159}$Dy\newline }$\varepsilon $: 100 \%\textbf{} & 144.4 d & 58.0 & 2.27 & ${}^{160}$Dy(p,pn)\newline ${}^{161}$Dy(p,p2n)\newline ${}^{162}$Dy(p,p3n)\newline ${}^{163}$Dy(p,p4n)\newline ${}^{164}$Dy(p,p5n)\newline ${}^{159}$Ho decay & ~-8575.9\newline -15030.29\newline -23227.29\newline ~-29498.3\newline ~-37156.42 \\ \hline 
\textbf{${}^{157}$Dy\newline }$\varepsilon $: 100 \%~\textbf{} & 8.14 h & 182.424\newline 326.336 & 1.33\newline 93 & ${}^{158}$Dy(p,pn)\newline ${}^{160}$Dy(p,p3n)\newline ${}^{161}$Dy(p,p4n)\newline ${}^{162}$Dy(p,p5n)\newline ${}^{163}$Dy(p,p6n)\newline ${}^{164}$Dy(p,p7n)\newline ${}^{157}$Ho decay & -9055.54\newline -24464.14\newline -30918.53\newline -39115.52\newline -45386.54\newline -53044.66 \\ \hline 
\textbf{${}^{155}$Dy\newline }$\varepsilon $: 98.62 \%\newline ${\beta}^{+}$:${}^{ }$1.38 \%~\textbf{} & 9.9 h & 184.564\newline 226.918 & 3.37\newline 68.4 & ${}^{158}$Dy(p,p3n)\newline ${}^{160}$Dy(p,p5n)\newline ${}^{161}$Dy(p,p6n)\newline ${}^{162}$Dy(p,p7n)\newline ${}^{163}$Dy(p,p8n)\newline ${}^{164}$Dy(p,p9n)\newline ${}^{155}$Ho decay & -25466.2\newline -40874.7\newline -47329.1\newline -55526.1\newline -61797.1\newline -69455.3 \\ \hline 
\textbf{${}^{161}$Tb\newline }$\beta $${}^{-}$: 100 \%~\textbf{} & 6.89 d & 74.56669\newline 87.941\newline 103.065\newline 106.113\newline 292.401 & 10.2\newline 0.183\newline 0.101\newline 0.078\newline 0.058 & ${}^{162}$Dy(p,2p)\newline ${}^{163}$Dy(p,2pn)\newline ${}^{164}$Dy(p,2p2n)\newline  & -8007.59\newline -14278.6\newline -21936.72 \\ \hline 
\textbf{${}^{160}$Tb\newline }$\beta $${}^{-}$:100 \%~\textbf{} & 72.3 d & 86.7877\newline 298.5783\newline 879.378\newline 966.166\newline 1177.954 & 13.2\newline 26.1\newline 30.1\newline 25.1\newline 14.9 & ${}^{161}$Dy(p,2p)\newline ${}^{162}$Dy(p,2pn)\newline ${}^{163}$Dy(p,2p2n)\newline ${}^{164}$Dy(p,2p3n) & -7507.17\newline -15704.16\newline -21975.17\newline -29633.29 \\ \hline 
\textbf{${}^{156}$Tb\newline }$\varepsilon $: 100 \%~\textbf{} & 5.35 d & 88.97\newline 199.19\newline 262.54\newline 296.49\newline 356.38\newline 422.34\newline 534.29\newline 1065.11\newline 1154.07\newline 1222.44 & 18\newline 41\newline 5.8\newline 4.5\newline 13.6\newline 8.0\newline 67\newline 10.8\newline 10.4\newline 31 & ${}^{158}$Dy(p,2pn)\newline ${}^{160}$Dy(p,2p3n)\newline ${}^{161}$Dy(p,2p4n)\newline ${}^{162}$Dy(p,2p5n)\newline ${}^{163}$Dy(p,2p6n)\newline ${}^{164}$Dy(p,2p7n) & -15674.87\newline -31083.47\newline -37537.86\newline -45734.85\newline -52005.87\newline -59663.98 \\ \hline 
\textbf{${}^{155}$Tb\newline }$\varepsilon $: 100 \%\textbf{} & 5.32 d & 86.55\newline 105.318\newline 148.64\newline 161.29\newline 163.28\newline 180.08\newline 262.27 & 32.0\newline 25.1\newline 2.65\newline 2.76\newline 4.44\newline 7.5\newline 5.3 & ${}^{158}$Dy(p,2p2n)\newline ${}^{160}$Dy(p,2p4n)\newline ${}^{161}$Dy(p,2p5n)\newline ${}^{162}$Dy(p,2p6n)\newline ${}^{163}$Dy(p,2p7n)\newline ${}^{164}$Dy(p,2p8n)\newline ${}^{155}$Dy decay & -22589.3\newline -37997.9\newline -44452.3\newline -52649.3\newline -58920.3\newline -66578.4 \\ \hline 
\end{tabular}

\end{center}
\end{table*}

\section{Results and discussion}
\label{3}

\subsection{Cross-sections}
\label{3.1}
The measured cross-sections for the production of $^{162m}Ho$, $^{161}Ho$, $^{159}Ho$, $^{159}Dy$, $^{157}Dy$, $^{155}Dy$, $^{161}Tb$, $^{160}Tb$, $^{156}Tb$, $^{155}Tb$ are shown in Table 3-4 and Figures 2-11. The figures also show the theoretical results calculated with the ALICE-IPPE and the EMPIRE codes and the values available in the on-line libraries TENDL-2011 and TENDL-2012 in comparison with experimental results of this work. We show both TENDL versions, obtained from the standard set - not adjusted parameters - to illustrate the difference and the effect of the upgrading. 
Due to the experimental circumstances (stacked foil technique, large dose at EOB, limited detector capacity) no cross-section data were obtained for short-lived activation products as $^{164m}Ho$(37.5 min), $^{164g}Ho$(29 min), $^{162g}Ho$(15.0 min), $^{160g}Ho$(25.6 min), $^{158m}Ho$(28 min), $^{157}Ho$(12.6 min), $^{156}Ho$ (9.5 s, 7.8 min, 56 min) and $^{155}Ho$, (48 min). 
Also the possibly produced radioisotopes $^{158}Tb$ (very long half-life, 180 a) and $^{160m}Ho$ (5.02 h) could not be identified in the measured spectra because of  low energy unresolved  $\gamma$-rays, or small effective cross-section due to the low abundance. 
The reactions are discussed separately for each reaction product. Naturally occurring dysprosium is composed of 7 stable isotopes ($^{156}Dy$ - 0.06\%, $^{158}Dy$ - 0.10\%, $^{160}Dy$ - 2.34\%, $^{161}Dy$ - 18.9\%, $^{162}Dy$ - 25.5\%, $^{163}Dy$ - 24.9\% and $^{164}Dy$ - 28.2\%). The relevant contributing reactions are collected in Table. 2.
The holmium reaction products are produced only through (p,xn) reactions, the dysprosium products directly via (p,pxn) reactions and through the decay of holmium radio-parents, the terbium radioisotopes are produced through directly (p,2pxn) reactions (including complex particle emissions) and decay of simultaneously produced dysprosium radio-products.

\subsubsection{$^{nat}Dy$(p,xn)$^{162m}Ho$ reaction}
\label{3.1.1}
Cross-sections for the short-lived $^{162g}Ho$ ground state ($T_{1/2}$ =15.0 min) were not measured. Theoretical estimates of $^{162m}Ho$ ($T_{1/2}$ =67.0 min) in TENDL 2011 and 2012 are systematically higher by a factor of 1.3 than our experimental results (Fig. 2). In case of ALICE-D and EMPIRE-D the agreement with the experiment is much better. To demonstrate the overestimation of TENDL, 0.7*TENDL-2012 results were also presented, which gives a good estimation in shape and value of the new experimental data.

\begin{figure}[h]
\includegraphics[scale=0.3]{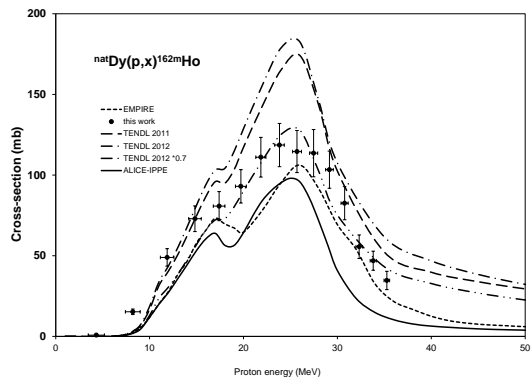}
\caption{Experimental and theoretical cross-sections for the formation of $^{162m}Ho$ by the proton bombardment of dysprosium}
\end{figure}

\subsubsection{$^{nat}Dy$(p,xn)$^{161}Ho$ reaction}
\label{3.1.2}
The experimental and theoretical cross-sections of the reactions producing $^{161}Ho$ ($T_{1/2}$ =2.48 h) are shown in Fig. 3.  The theoretical overestimation in all cases is significant. Scaling by a factor 0.7 of the TENDL-2011 results shows a rather good agreement. 

\begin{figure}[h]
\includegraphics[scale=0.3]{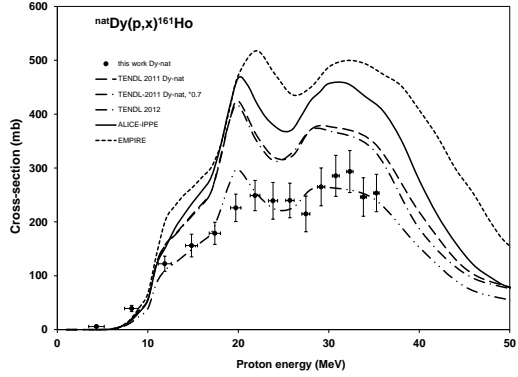}
\caption{Experimental and theoretical cross-sections for the formation of $^{161}Ho$ by the proton bombardment of dysprosium}
\end{figure}

\subsubsection{$^{nat}Dy$(p,xn)$^{159}Ho$ reaction}
\label{3.1.3}
The theory follows both in shape and in magnitude the experimental cross-sections of the $^{159}Ho$ ($T_{1/2}$ = 33.05 min) in the investigated energy range (Fig. 4), as far as the TENDL calculations regarded. The ALICE-IPPE and EMPIRE overestimate again.

\begin{figure}[h]
\includegraphics[scale=0.3]{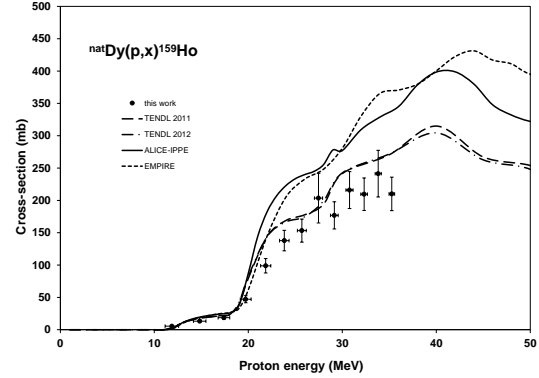}
\caption{Experimental and theoretical cross-sections for the formation of $^{159}Ho$ by the proton bombardment of dysprosium}
\end{figure}

\subsubsection{P$^{nat}Dy$(p,x)$^{159}Dy$ reaction}
\label{3.1.4}
The cumulative cross-sections of reactions producing $^{159}Dy$ ($T_{1/2}$ = 144.4 d) contain apart from the direct production, the contribution from the decay of $^{159}Ho$ ($T_{1/2}$ = 33.05 min) as they were measured after nearly complete decay of the parent isotope. The agreement with the results of the 3 codes is acceptable (Fig. 5). The TENDL results show that the direct production is negligible, especially below 30 MeV.

\begin{figure}[h]
\includegraphics[scale=0.3]{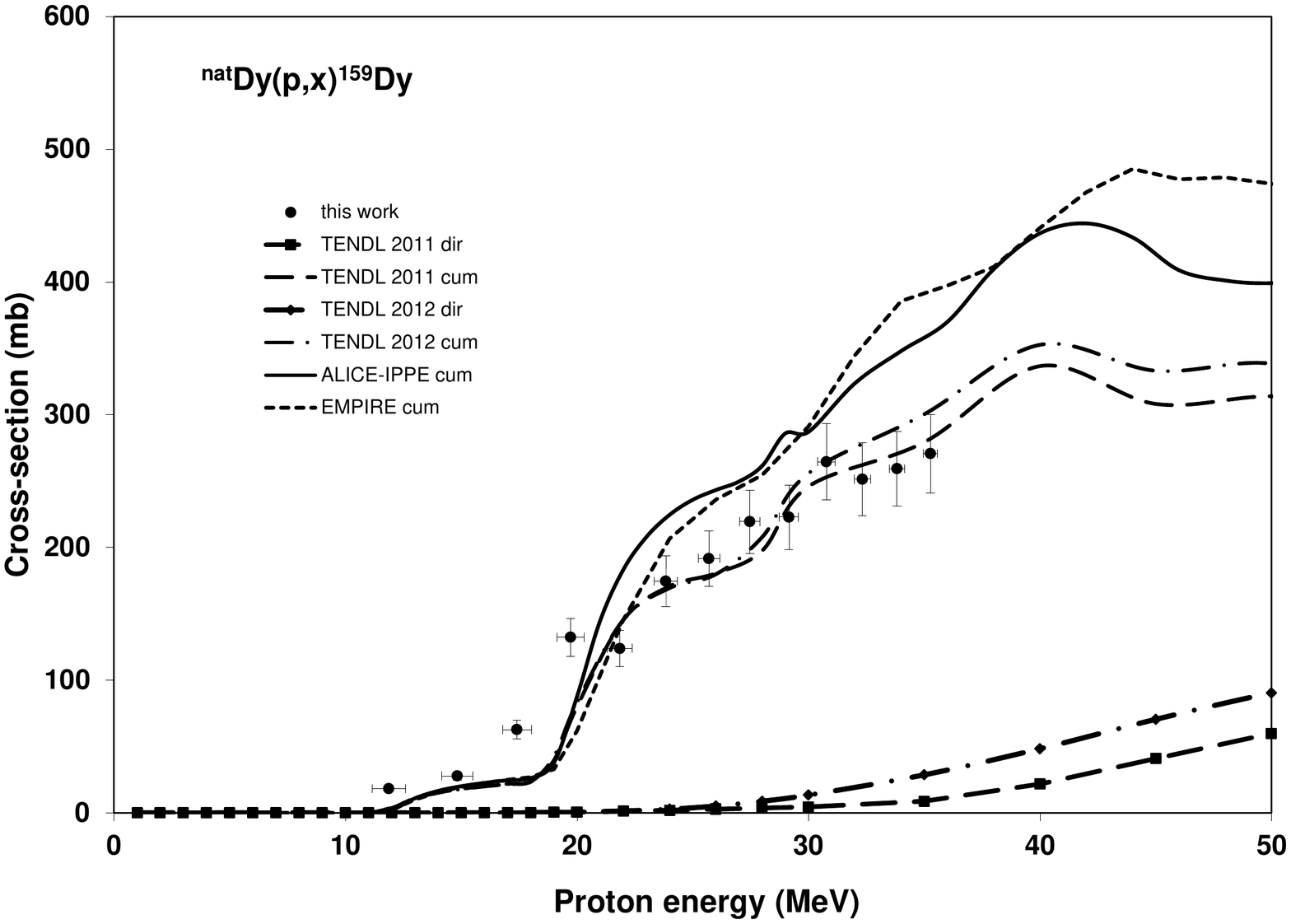}
\caption{Experimental and theoretical cross-sections for the formation of $^{159}Dy$ by the proton bombardment of dysprosium}
\end{figure}

\subsubsection{$^{nat}Dy$(p,x)$^{157}Dy$ reaction}
\label{3.1.5}
The cumulative cross-sections for production of $^{157}Dy$ ($T_{1/2}$ =8.14 h) were measured after nearly complete decay of the parent $^{157}Ho$ ($T_{1/2}$ =12.6 min). The theoretical data overestimate the experimental results and here also the TENDL results show that the direct production is negligible (Fig. 6).

\begin{figure}[h]
\includegraphics[scale=0.3]{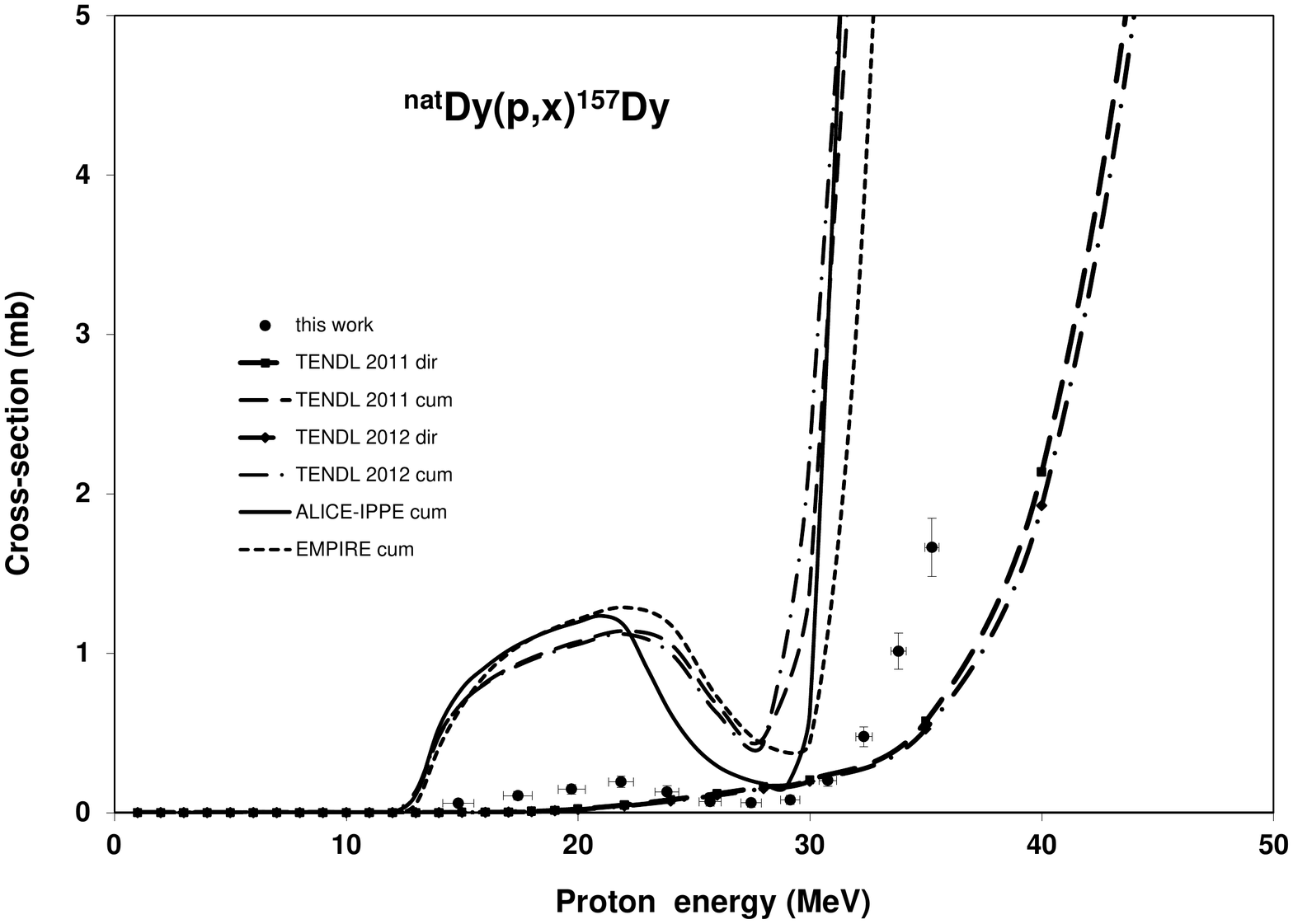}
\caption{Experimental and theoretical cross-sections for the formation of $^{157}Dy$ by the proton bombardment of dysprosium}
\end{figure}

\subsubsection{$^{nat}Dy$(p,x)$^{155}Dy$ reaction}
\label{3.1.6}
The measured $^{155}Dy$ ($T_{1/2}$ = 9.9 h) was produced directly and through decay of the $^{155}Ho$ ($T_{1/2}$ = 48 min) parent radioisotope. The comparison with the TENDL results shows good agreement below 30 MeV (Fig. 7), and the agreement with the ALICE-IPPE and the EMPIRE results is also good below 25 MeV.

\begin{figure}[h]
\includegraphics[scale=0.3]{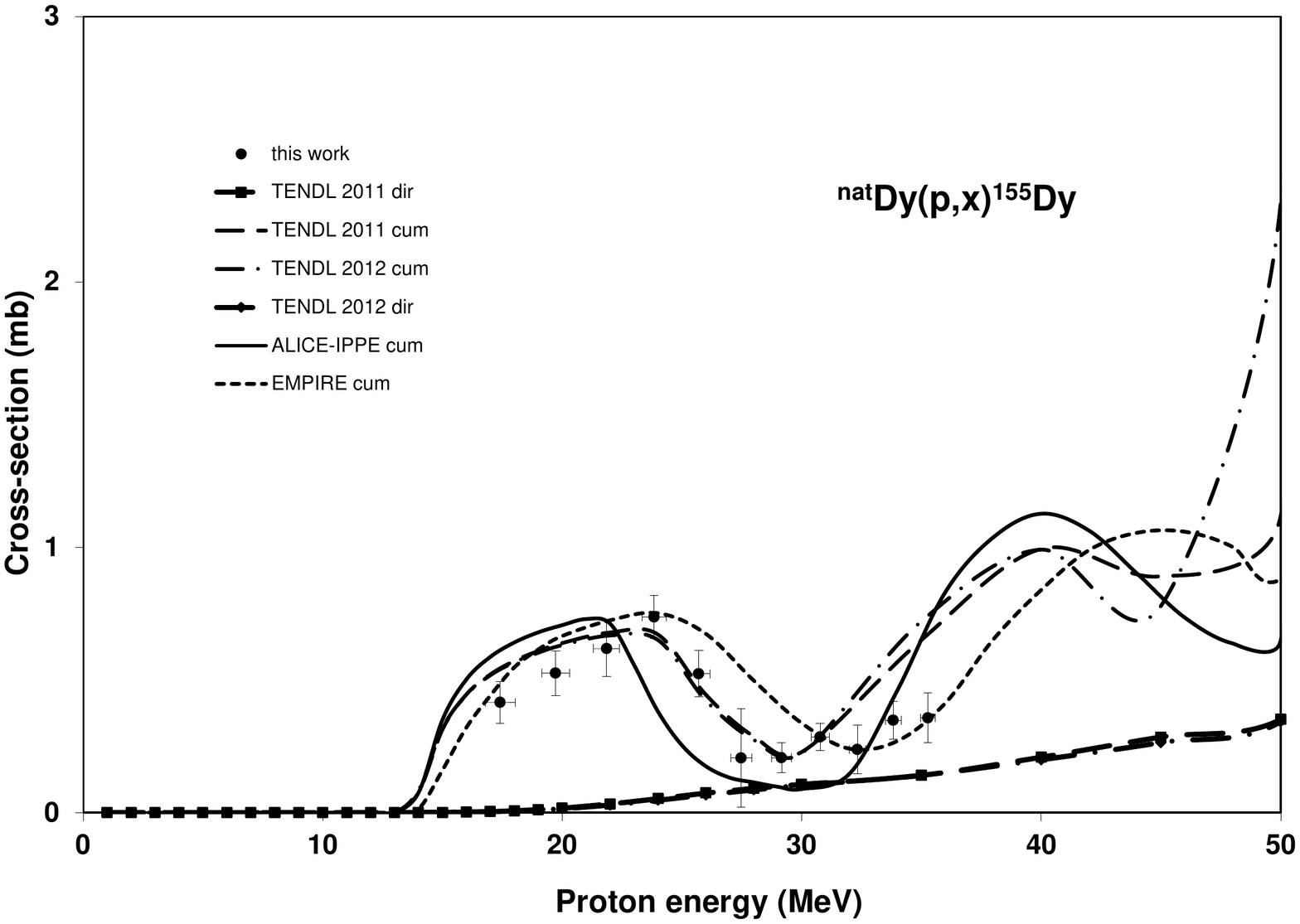}
\caption{Experimental and theoretical cross-sections for the formation of $^{155}Dy$ by the proton bombardment of dysprosium}
\end{figure}

\subsubsection{$^{nat}Dy$(p,x)$^{161}Tb$ reaction}
\label{3.1.7}
The measured direct cross-sections for production of $^{161}Tb$($T_{1/2}$ = 6.89 d) are shown in Fig. 8. As can be seen in Table 2 and 3, reactions on stable Dy target isotopes with nearly the same abundance can contribute to the production of $^{161}Tb$. From systematics it is known that the (p,2p) reaction has mostly lower cross-sections than the (p,2pn) channel. The sharp low energy peak due to $^{164}Dy$(p, p) that can be seen in the TENDL theoretical data is hence questionable and is not reproduced by the experimental values, neither by the EMPIRE nor by ALICE results. It has to be mentioned too that the three codes provide very different results both in shape and in magnitude. 

\begin{figure}[h]
\includegraphics[scale=0.3]{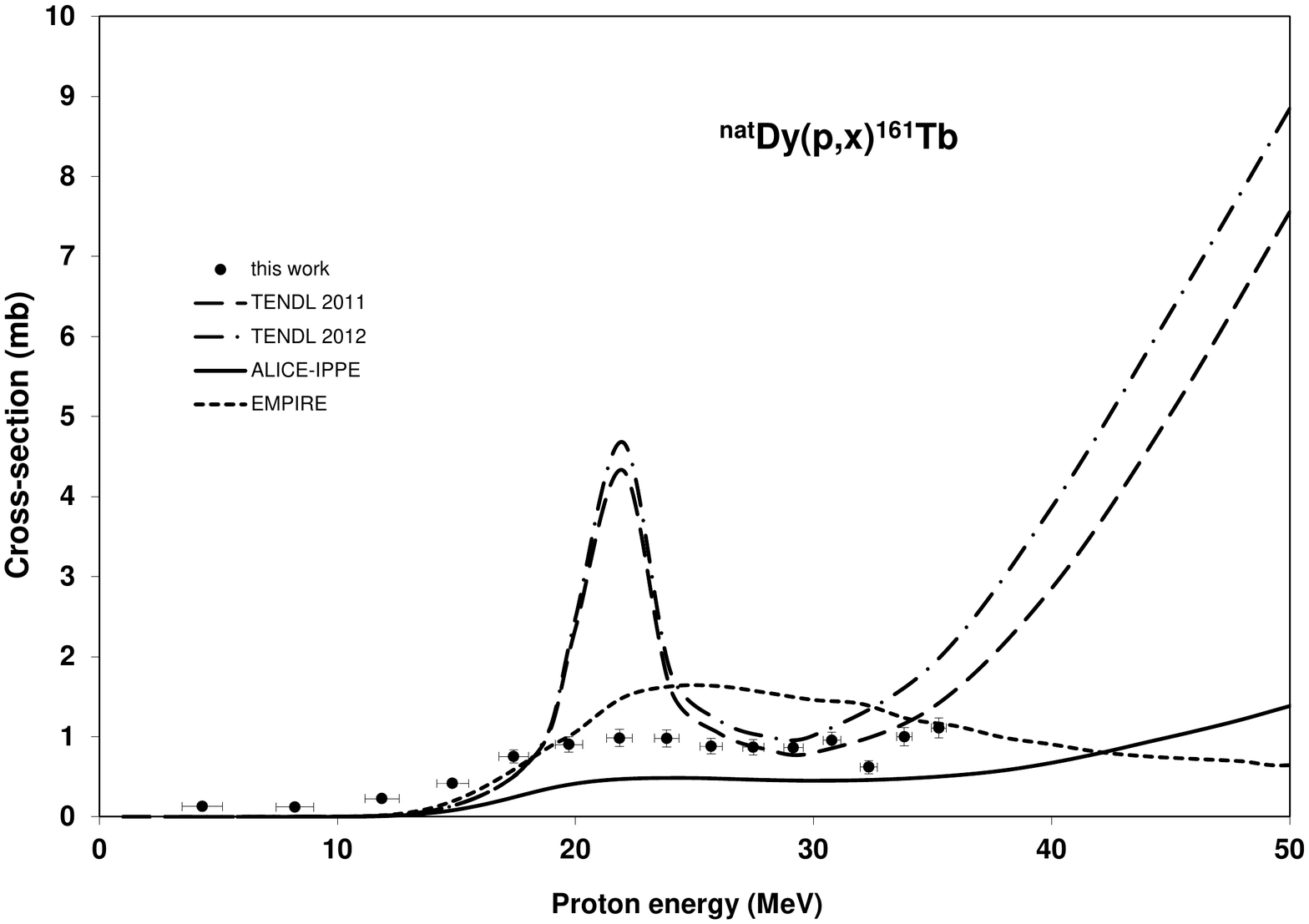}
\caption{Experimental and theoretical cross-sections for the formation of $^{161}Tb$ by the proton bombardment of dysprosium}
\end{figure}

\subsubsection{n$^{nat}Dy$(p,xn)$^{160}Tb$ reaction}
\label{3.1.8}
Here also a sharp peak is predicted by the TENDL (Fig. 9) around 20 MeV, but according to physics expectation, the results of other model codes and to the experiment, no such peak is present in the excitation function of the $^{160}Tb$ ($T_{1/2}$ = 72.3 d) in the investigated energy range. EMPIRE and ALICE describe well the shape but overestimate, respectively underestimate by a factor of two the experimental values.

\begin{figure}[h]
\includegraphics[scale=0.3]{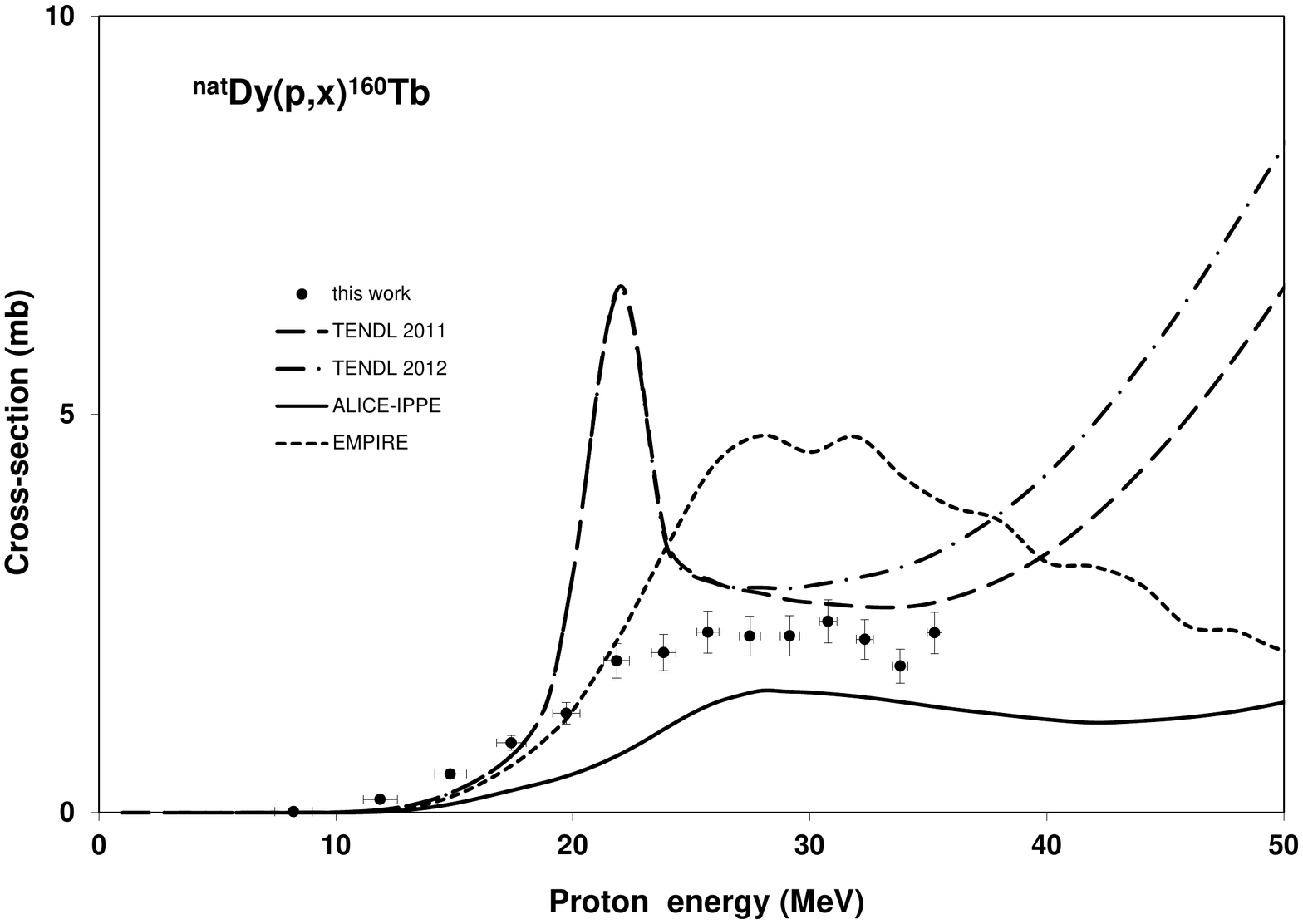}
\caption{Experimental and theoretical cross-sections for the formation of $^{160}Tb$ by the proton bombardment of dysprosium}
\end{figure}

\subsubsection{$^{nat}Dy$(p,xn)$^{156}Tb$ reaction}
\label{3.1.9}
The cross-sections of $^{156g}Tb$ ($T_{1/2}$ =5.35 d) were obtained from spectra measured after complete decay of the 2 isomeric states ($T_{1/2}$ =5.3 h, IT: 100 \% and $T_{1/2}$ = 24.4 h, IT: 100 \%). In the investigated energy range there is a good resemblance between the data from ALICE the experimental data (Fig. 10). In this case TENDL-2011 and TENDL-2012 do not present a peaked structure but underestimate the experimental cross-sections while EMPIRE again overestimates by a factor of 3.

\begin{figure}[h]
\includegraphics[scale=0.3]{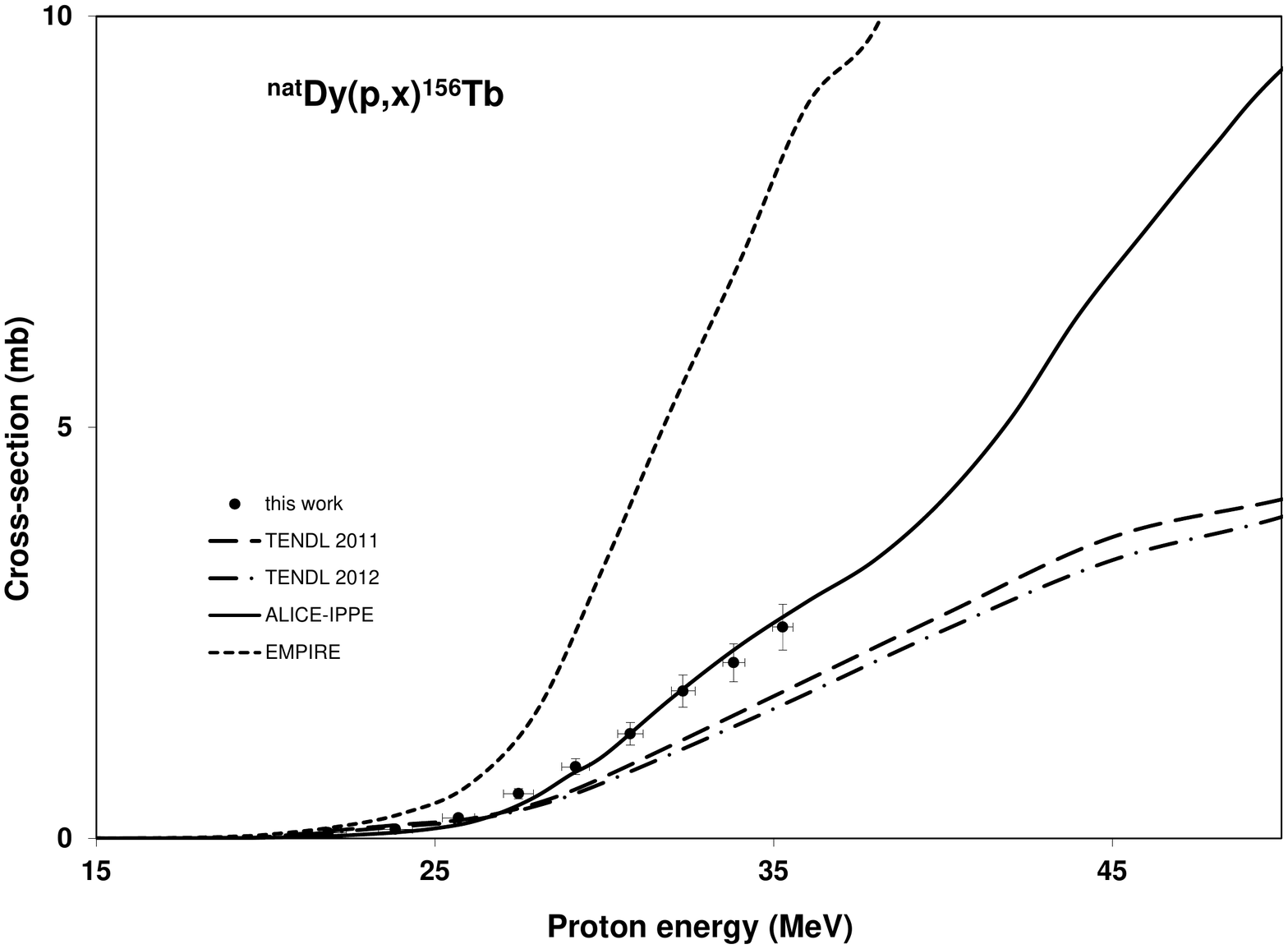}
\caption{Experimental and theoretical cross-sections for the formation of $^{156}Tb$ by the proton bombardment of dysprosium}
\end{figure}

\subsubsection{$^{nat}Dy$(p,xn)$^{155}Tb$ reaction}
\label{3.1.10}
The measured cross-sections of $^{155}Tb$ ($T_{1/2}$ = 5.32 d) contains the complete contribution from the decay of $^{155}Dy$ ($T_{1/2}$ = 9.9 h). The theories predict cross-sections close to the experimental data in the investigated energy range (Fig. 11).

\begin{figure}[h]
\includegraphics[scale=0.3]{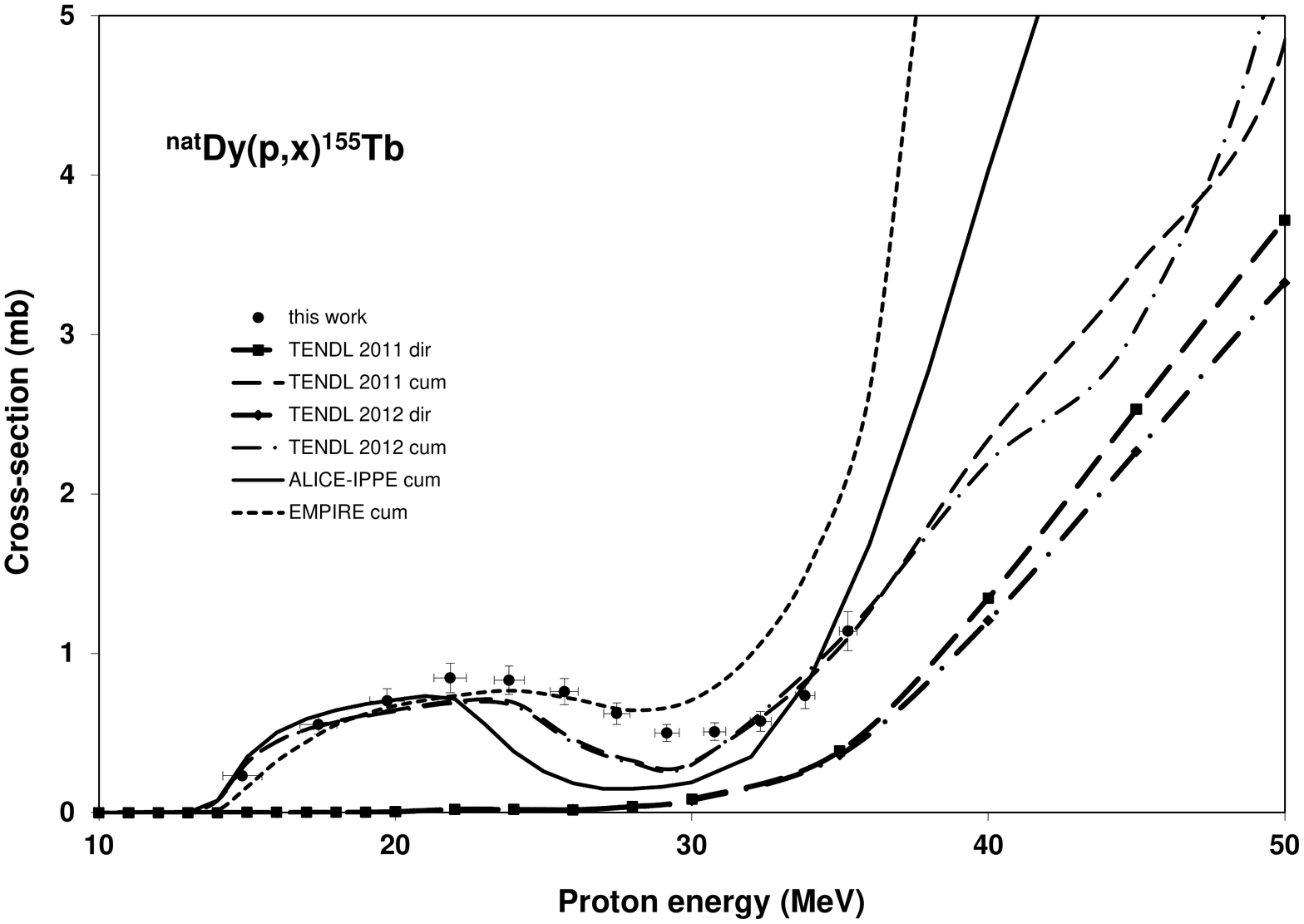}
\caption{Experimental and theoretical cross-sections for the formation of $^{155}Tb$ by the proton bombardment of dysprosium}
\end{figure}

\begin{table*}[t]
\tiny
\caption{Experimental cross-sections of ${}^{nat}$Dy(p,xn)${}^{ }$${}^{162m}$Ho, ${}^{161}$Ho, ${}^{159}$Ho, ${}^{159}$Dy,${}^{ 157}$Dy,${}^{ 155}$Dy reactions }
\centering
\begin{center}
\begin{tabular}{|p{0.3in}|p{0.3in}|p{0.3in}|p{0.3in}|p{0.3in}|p{0.3in}|p{0.3in}|p{0.3in}|p{0.3in}|p{0.3in}|p{0.3in}|p{0.3in}|p{0.3in}|p{0.3in}|} \hline 
\multicolumn{2}{|c|}{\textbf{E $\pm\Delta$E \newline (MeV)}} & \multicolumn{12}{|c|}{\textbf{Cross-section($\sigma$)$\pm\Delta\sigma$ \newline (mb)}} \\ \hline 
\multicolumn{2}{|p{1in}|}{\textbf{}} & \multicolumn{2}{|p{0.7in}|}{\textbf{${}^{162m}$Ho}} & \multicolumn{2}{|p{0.5in}|}{\textbf{${}^{161}$Ho}} & \multicolumn{2}{|p{0.7in}|}{\textbf{${}^{159}$Ho}} & \multicolumn{2}{|p{0.7in}|}{\textbf{${}^{159}$Dy}} & \multicolumn{2}{|p{0.7in}|}{\textbf{${}^{157}$Dy}} & \multicolumn{2}{|p{0.6in}|}{\textbf{${}^{155}$Dy}} \\ \hline 
35.3 & 0.3 & 34.7 & 5.7 & 253.5 & 34.8 & 210.0 & 25.9 & 270.7 & 29.4 & 1.66 & 0.18 & 0.36 & 0.09 \\ \hline 
33.8 & 0.3 & 46.9 & 5.9 & 246.2 & 35.6 & 241.4 & 36.1 & 259.3 & 28.2 & 1.01 & 0.11 & 0.35 & 0.07 \\ \hline 
32.3 & 0.4 & 55.6 & 7.3 & 293.4 & 39.0 & 209.6 & 25.2 & 251.4 & 27.4 & 0.48 & 0.06 & 0.24 & 0.09 \\ \hline 
30.8 & 0.4 & 82.6 & 10.2 & 285.5 & 38.1 & 216.0 & 28.6 & 264.5 & 28.8 & 0.20 & 0.04 & 0.28 & 0.05 \\ \hline 
29.2 & 0.4 & 103.4 & 11.5 & 265.0 & 35.2 & 176.8 & 21.1 & 222.8 & 24.3 & 0.08 & 0.03 & 0.21 & 0.06 \\ \hline 
27.5 & 0.4 & 113.6 & 14.6 & 214.8 & 33.2 & 203.5 & 38.4 & 219.4 & 23.9 & 0.06 & 0.03 & 0.21 & 0.19 \\ \hline 
25.7 & 0.5 & 114.6 & 13.1 & 239.8 & 32.2 & 153.3 & 17.9 & 191.6 & 20.9 & 0.07 & 0.03 & 0.52 & 0.09 \\ \hline 
23.8 & 0.5 & 118.6 & 13.4 & 239.2 & 34.1 & 137.8 & 15.8 & 174.5 & 19.1 & 0.13 & 0.04 & 0.74 & 0.08 \\ \hline 
21.9 & 0.5 & 111.1 & 12.3 & 248.9 & 28.1 & 98.9 & 11.2 & 123.8 & 13.6 & 0.19 & 0.03 & 0.62 & 0.10 \\ \hline 
19.7 & 0.6 & 93.0 & 10.4 & 226.1 & 25.5 & 47.1 & 5.5 & 132.2 & 14.4 & 0.15 & 0.03 & 0.53 & 0.08 \\ \hline 
17.4 & 0.6 & 80.8 & 9.0 & 178.8 & 20.6 & 18.7 & 2.5 & 62.7 & 6.8 & 0.11 & 0.03 & 0.42 & 0.08 \\ \hline 
14.8 & 0.7 & 73.0 & 8.0 & 156.1 & 21.1 & 13.3 & 1.7 & 27.7 & 3.0 & 0.06 & 0.02 &  & ~ \\ \hline 
11.9 & 0.7 & 49.0 & 5.4 & 122.3 & 14.1 & 5.5 & 0.7 & 18.2 & 2.0 &  & ~ &  & ~ \\ \hline 
8.2 & 0.8 & 15.3 & 1.7 & 39.2 & 5.8 &  & ~ &  & ~ &  & ~ &  & ~ \\ \hline 
4.3 & 0.9 & 0.8 & 0.1 & 5.7 & 1.7 &  & ~ &  & ~ &  & ~ &  & ~ \\ \hline 
\end{tabular}

\end{center}
\end{table*}

\begin{table*}[t]
\tiny
\caption{Experimental cross-sections  of ${}^{nat}$Dy(p,xn)${}^{ }$${}^{161}$Tb, ${}^{160}$Tb, ${}^{156}$Tb and${}^{ 155}$Tb reactions }
\centering
\begin{center}
\begin{tabular}{|p{0.3in}|p{0.3in}|p{0.3in}|p{0.3in}|p{0.3in}|p{0.3in}|p{0.3in}|p{0.3in}|p{0.3in}|p{0.3in}|} \hline 
\multicolumn{2}{|c|}{\textbf{E $\pm\Delta$E \newline (MeV)}} & \multicolumn{8}{|c|}{\textbf{Cross-section($\sigma$)$\pm\Delta\sigma$ \newline (mb)}} \\ \hline 
\multicolumn{2}{|p{1in}|}{} & \multicolumn{2}{|p{0.7in}|}{\textbf{${}^{161}$Tb}} & \multicolumn{2}{|p{0.5in}|}{\textbf{${}^{160}$Tb}} & \multicolumn{2}{|p{0.7in}|}{\textbf{${}^{156}$Tb}} & \multicolumn{2}{|p{0.7in}|}{\textbf{${}^{155}$Tb}} \\ \hline 
35.3 & 0.3 & 1.11 & 0.12 & 2.26 & 0.26 & 2.57 & 0.28 & 1.14 & 0.12 \\ \hline 
33.8 & 0.3 & 1.00 & 0.11 & 1.84 & 0.21 & 2.14 & 0.23 & 0.73 & 0.08 \\ \hline 
32.3 & 0.4 & 0.62 & 0.08 & 2.18 & 0.25 & 1.79 & 0.19 & 0.57 & 0.06 \\ \hline 
30.8 & 0.4 & 0.96 & 0.11 & 2.40 & 0.27 & 1.27 & 0.14 & 0.51 & 0.06 \\ \hline 
29.2 & 0.4 & 0.86 & 0.10 & 2.22 & 0.25 & 0.87 & 0.10 & 0.50 & 0.06 \\ \hline 
27.5 & 0.4 & 0.87 & 0.10 & 2.22 & 0.25 & 0.54 & 0.06 & 0.62 & 0.07 \\ \hline 
25.7 & 0.5 & 0.88 & 0.10 & 2.27 & 0.26 & 0.25 & 0.03 & 0.76 & 0.08 \\ \hline 
23.8 & 0.5 & 0.98 & 0.11 & 2.01 & 0.23 & 0.11 & 0.01 & 0.83 & 0.09 \\ \hline 
21.9 & 0.5 & 0.98 & 0.11 & 1.91 & 0.22 & 0.07 & 0.01 & 0.85 & 0.09 \\ \hline 
19.7 & 0.6 & 0.90 & 0.10 & 1.25 & 0.14 &  & ~ & 0.70 & 0.08 \\ \hline 
17.4 & 0.6 & 0.75 & 0.08 & 0.88 & 0.10 &  & ~ & 0.55 & 0.06 \\ \hline 
14.8 & 0.7 & 0.42 & 0.05 & 0.49 & 0.06 &  & ~ & 0.23 & 0.03 \\ \hline 
11.9 & 0.7 & 0.23 & 0.03 & 0.17 & 0.02 &  & ~ &  & ~ \\ \hline 
8.2 & 0.8 & 0.12 & 0.01 & 0.01 & 0.01 &  & ~ &  & ~ \\ \hline 
4.3 & 0.9 & 0.13 & 0.01 &  & ~ &  & ~ &  & ~ \\ \hline 
\end{tabular}

\end{center}
\end{table*}

\subsection{Integral yields}
\label{4.}
The integral yields calculated from spline fits to our experimental excitation functions are shown in Fig 12 and Fig. 13. The integral yields represent so called physical yields i.e. activity instantaneous production rates at 1 $\mu$A beam current \citep{Bonardi}. The only literature value found is for $^{161}Ho$ \citep{Stephens2010}, which is about 30\% lower than our calculated curve. From Fig. 12. It is seen that the produced radio-isotopes form two groups from the point of view of production yield. The production of holmium radio-isotopes begins at much lower bombarding energies and the production yields are also 3-4 orders of magnitude higher, that's why this figure was plotted with logarithmic scale. In Fig. 13 the terbium radio-isotopes are presented. The most promising is the $^{161}Tb$ both from the point of view of production yield and threshold of the production.

\begin{figure}[h]
\includegraphics[scale=0.3]{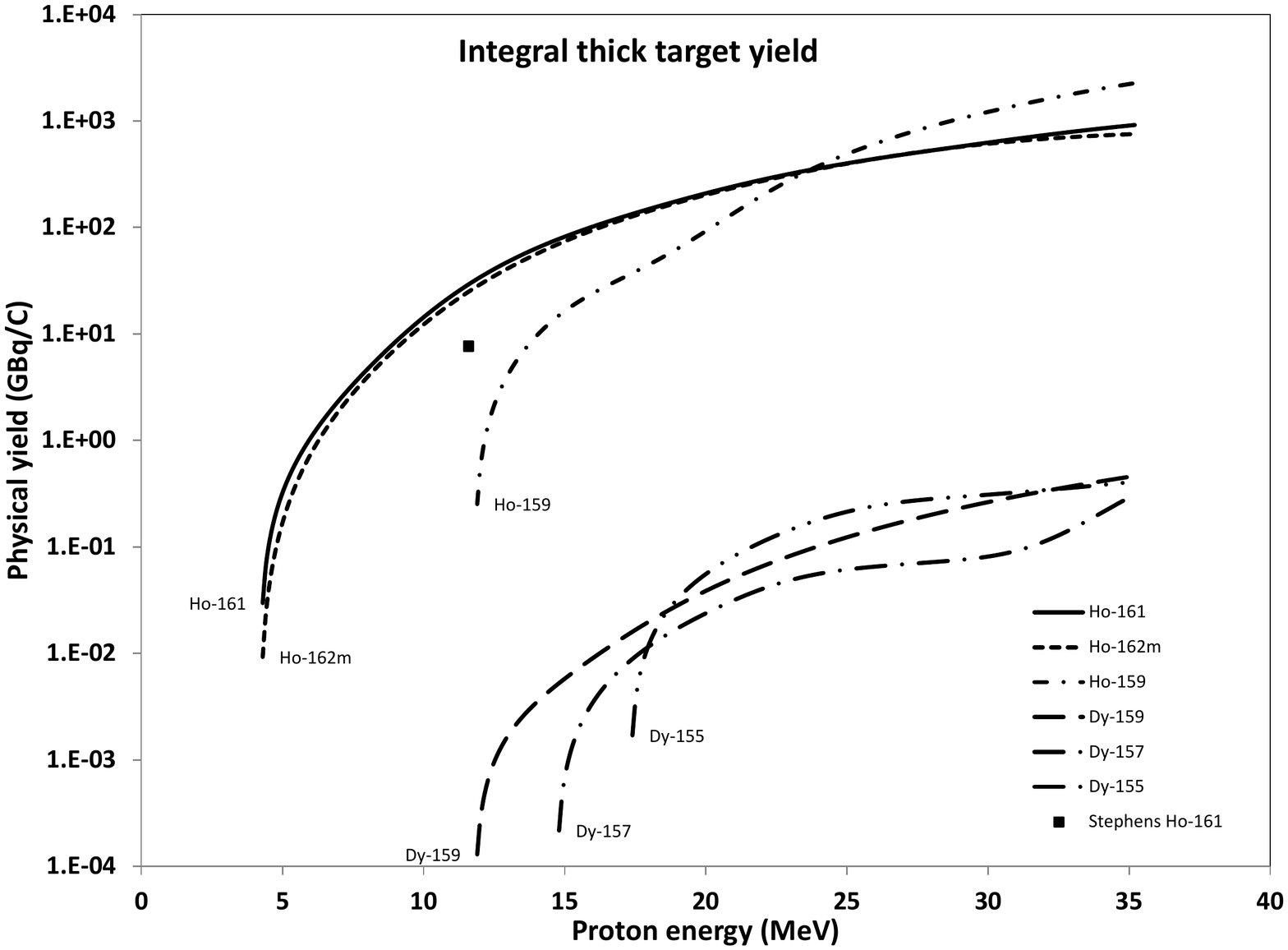}
\caption{Integral thick target yields for the formation of $^{162m}Ho$, $^{161}Ho$, $^{159}Ho$, $^{159}Dy$, $^{157}Dy$, $^{155}Dy$ in proton induced nuclear reaction on $^{nat}Dy$ as a function of the energy}
\end{figure}

\begin{figure}[h]
\includegraphics[scale=0.3]{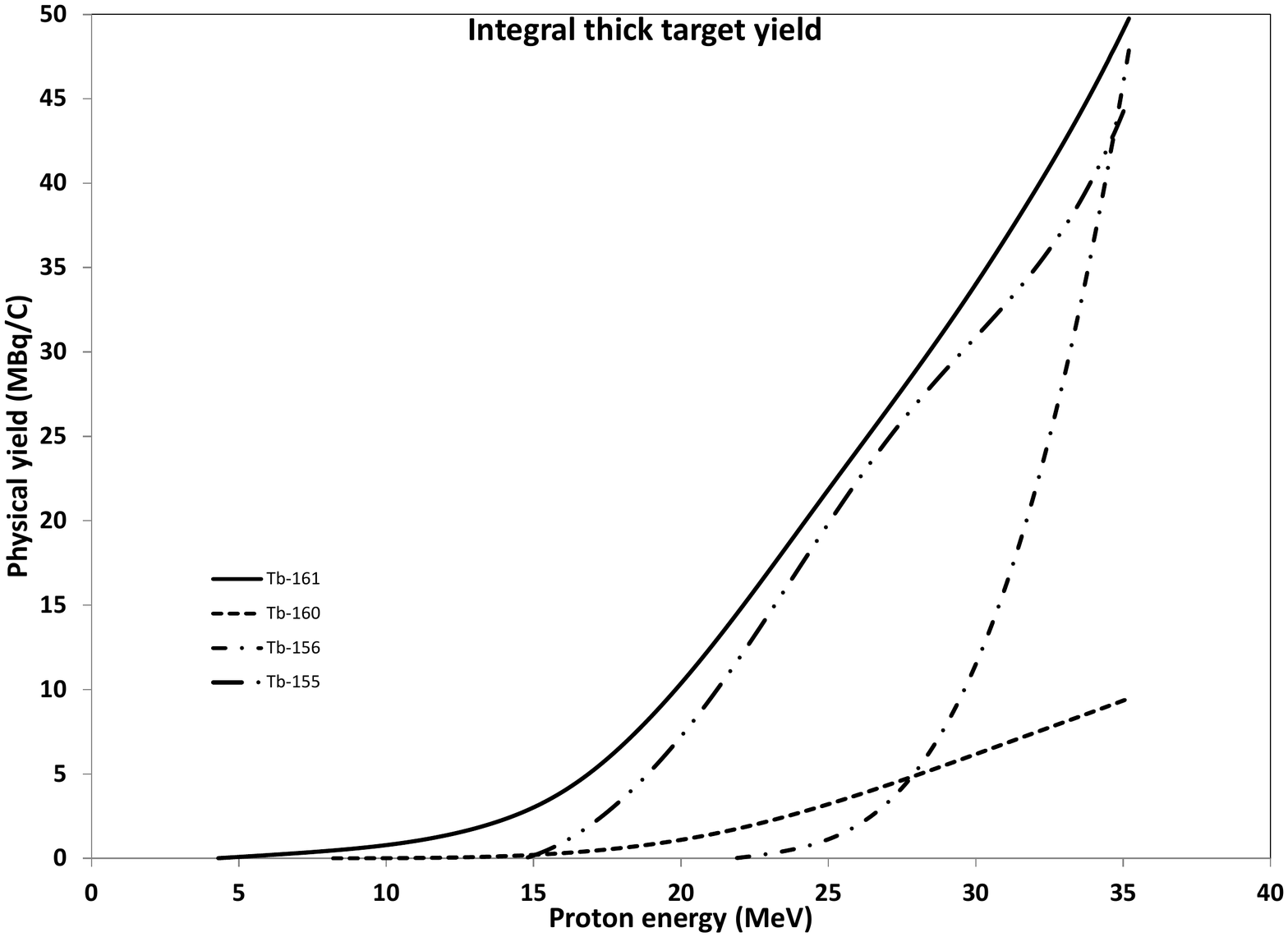}
\caption{Integral thick target yields for the formation of $^{161}Tb$, $^{160}Tb$, $^{156}Tb$, $^{155}Tb$ in proton induced nuclear reaction on $^{nat}Dy$ as a function of the energy}
\end{figure}

\section{Summary and applications}
\label{5}
We present the first experimental cross-sections for the $^{nat}Dy$(p,x) $^{162m}Ho$, $^{161}Ho$, $^{159}Ho$, $^{159}Dy$, $^{157}Dy$, $^{155}Dy$, $^{161}Tb$, $^{160}Tb$, $^{156}Tb$ and $^{155}Tb$ nuclear reactions up to 36 MeV incident proton energy. Our excitation functions are relative to the $^{nat}Ti$(p,n)$^{48}V$ monitor reaction measured simultaneously. The experimental data were compared with the results obtained by the TALYS code reported in the TENDL libraries and with the results of ALICE-IPPE and EMPIRE calculations. The theoretical description, even for these proton induced reactions, is disappointing and gives disagreements.
While for the (p,xn) reactions, resulting in Ho radionuclides, the predicted shapes are rather well corresponding to the contributions on multiple stable target nuclides, the absolute values are sometimes too high, sometimes too low by 50 to 100\% and without a systematic behavior of each code.
For formation of Dy, reactions with emission of a proton and decay of parent radionuclides are involved, the agreement between the different codes and the experimental values is better. The discrepancies are probably due to poor description of the parent Ho radionuclides cross-sections as the direct reaction cross-sections are representing less than 10 \% of the cumulative production.
For Tb radionuclides in some cases the calculated excitation functions are well representing the experimental results but again disagreements in amplitude and non-physical peaks are appearing with no systematic behavior for "good" or "bad" codes.
The differences between the two versions of the TENDL libraries are minimal and appear mostly as differences in amplitude of the order of 10\% and especially at energies above 30 MeV.
The experimental data continue hence to be important for testing the predictivity and improving the performances of the model codes, taking into account that no other experimental activation data are available for these reactions. 
The experimental results are of importance for several practical applications. Among the investigated reaction products we can mention: 
\begin{itemize} 
\item	the radio-lanthanide $^{161}Ho$ ($T_{1/2}$ =6.9 d) is a promising Auger-electron emitter for internal radiotherapy, \citep{Neves, Rosch, Stephens2010, Stephens2010b,UUsijarvi}
\item	The radionuclide $^{159}Dy$ ($T_{1/2}$ = 144 d, EC =100\%) is a pure Auger electron and X-ray emitter and has gained interest in transmission imaging and bone mineral analysis \citep{Nayak}, while$^{157}Dy$ ($T_{1/2}$ =8.14 h.)\citep{Lebowitz}, was investigated as a bone seekers in the evaluation of bone lesions \citep{Hubner}.
\end{itemize}

Terbium offers 4 clinically interesting radioisotopes with complementary physical decay characteristics: $^{149}Tb$, $^{152}Tb$, $^{155}Tb$, and $^{161}Tb$. The identical chemical characteristics of these radioisotopes allow the preparation of radiopharmaceuticals with identical pharmacokinetics useful for PET ($^{152}Tb$) or SPECT diagnosis ($^{155}Tb$) and for ?- ($^{149}Tb$) and $\beta^{-}$-particle ($^{161}Tb$) therapy \citep{Muller}.  From Dy targets the higher mass $^{155}Tb$ (decay of $^{155}Dy$) and $^{161}Tb$ can be obtained, but we showed in a recent publication that other production methods are better \citep{TF20132}. Production routes on other elemental targets have to be considered for $^{149,152}Tb$.

\section{Acknowledgements}
\label{6}
This work was performed in the frame of the HAS-FWO Vlaanderen (Hungary-Belgium) project. The authors acknowledge the support of the research project and of the respective institutions. 
 


%
%

\clearpage
\bibliographystyle{elsarticle-harv}
\bibliography{Dyp}




%



\end{document}